\documentclass[10pt, conference]{IEEEtran}
\usepackage{epsfig,color,amsmath,cite}
\usepackage{amsthm}
\usepackage{wrapfig}
\usepackage{bm}
\usepackage{epstopdf}
\usepackage{amssymb}
\usepackage{url}
\usepackage{enumitem}
\usepackage{multirow}
\usepackage{booktabs}
\usepackage{tikz}
\usepackage{pgfplots}
\usepackage[linesnumbered,lined,boxed,commentsnumbered,ruled,longend]{algorithm2e}


\setlist[itemize]{leftmargin=*}

\makeatother

\allowdisplaybreaks[4]
\usepackage[colorlinks = true,
linkcolor = blue,
urlcolor  = blue,
citecolor = blue,
anchorcolor = blue]{}

\newcommand{\RomanNumeralCaps}[1]
{\MakeUppercase{\romannumeral #1}}

\usepackage{graphicx}
\usepackage{color}
\usepackage{amsfonts}
\usepackage{float}
\usepackage{amsmath}
\usepackage{amsthm}
\usepackage{epstopdf}
\usepackage{changepage}
\usepackage{latexsym}
\usepackage{tikz}
\usepackage{cases}
\usepackage{longtable}
\usepackage{supertabular}
\usepackage{tabu}
\usepackage{multirow}
\usepackage{multicol}
\usepackage{booktabs}
\usepackage{arydshln}
\usepackage{setspace}

\usepackage[noabbrev]{cleveref}

\usepackage{mathtools}

\DeclarePairedDelimiter\abs{\lvert}{\rvert}%
\DeclarePairedDelimiter\norm{\lVert}{\rVert}%

\makeatletter
\let\oldabs\abs
\def\abs{\@ifstar{\oldabs}{\oldabs*}}
\let\oldnorm\norm
\def\norm{\@ifstar{\oldnorm}{\oldnorm*}}
\makeatother

\DeclareMathAlphabet\mathbfcal{OMS}{cmsy}{b}{n}

\usepackage{stackengine}

\usepackage{graphicx}
\usepackage{float}
\usepackage[caption = false]{subfig}

\usepackage[labelfont=bf]{caption}

\IEEEoverridecommandlockouts
	\title{\vspace{0.4cm} \LARGE  \textbf{Impact of Distributed Energy Resources on Frequency Regulation of the Bulk Power System}}

	\author{Mohammad Khatibi and Sara Ahmed\\Department of Electrical and 	Computer Engineering\\University of Texas at San Antonio\\San Antonio, TX\\Mohammad.Khatibi@utsa.edu}

\begin{document}
	\maketitle
	
	\begin{abstract}
	
	The growing penetration of distributed energy resources (DERs) has increased the complexity of the power system due to their intermittent characteristics and lower inertial response, such as photo voltaic (PV) systems and wind turbines. This restructuring of power system has considerable effect on transient response of the system resulting in inter area oscillations, less synchronized coupling and power swings. Furthermore the concept of being distributed itself and generating electricity from multiple locations in power system makes the transient impact of DERs even worse by raising issues such as reverse power flows.
	This paper studies some impacts of the changing nature of power system which are limiting the large scale integration of DERs. In addition a solution to increase the inertial response of the system is addressed by adding virtual inertia to the inverter based DERs in power system.The proposed control results in increasing the stability margin and tracking the rated frequency of the system. The injected synchronized active power to the system will prevent the protection relays from tripping by improving the rate of change of frequency. The proposed system operation is implemented on a sample power grid comprising of generation, transmission and distribution and results are verified experimentally through the Opal-RT real-time simulation system.
	\end{abstract}

	\begin{IEEEkeywords}
	power system stability,distributed energy resources, photo voltaic system, virtual inertia, rate of change of frequency.
	\end{IEEEkeywords}

	\section{Introduction}

\cite{fang2017small}In recent years, significant inverter-based inertia-less renewable generation has been integrated in both bulk transmission and distribution (T\&D) power systems to improve the sustainability of electric power systems. The increasing penetration of the distributed energy resources (DERs) displacing conventional synchronous generators (SGs) is rapidly changing the dynamics of the large-scale power systems. The electric grids lose inertia, voltage support and oscillation damping. 
When majority of generated electricity is coming from synchronous generators working with fossil fuels, using DERs reduces the system fuel costs significantly but can have considerable impact on system reliability. This less reliable grid pushed the power system planners to develop a method that helps to decide on operating policies, mixes and sizes in capacity expansion and installation sites when utilizing wind and photo voltaic (PV) systems \cite{karki2001reliability}.
Network expansion planning, voltage stability studies and coordination of voltage controls at the T\&D interface are investigated through power flow which is from transmission to distribution in traditional power grids. Reverse power flow from DGs to transmission system and impact of DERs on voltage stability in  restructured power grids with high penetration of DERs asks for a new modeling and representation for DGs. For example in \cite{nikkhajoei2006steady,chen2006power} only positive sequence representation has been considered for power flow analysis in presence of DERs which is not enough for an unbalanced distribution grid with unbalanced laterals.
in \cite{nikkhajoei2006steady} a three-phase power-flow algorithm is proposed which includes unbalanced lines and loads, single phase laterals and three/four wire distribution lines. 
A detailed analysis of the impact of large scale wind power generation on the dynamic voltage stability and the transient stability of electric power systems is presented in \cite{slootweg2003impact}.
Using a multidimensional parameter variation \cite{dierkes2014impact} shows that different control strategies of renewable energy sources have a significant influence on voltage stability of the power system.
To verify the relay protection settings and operation and circuit breaker and fuse ratings, short circuit analysis should to be taken in into account. The dynamics of DERs should be included in short circuit analysis during fault in distribution system since each DER will contribute to short circuit current. 


 Inverter based DGs has no inertia. To solve this problem the idea of virtual synchronous machine/ generator (VSM/VSG) has been presented in \cite{tamrakar2017virtual,bevrani2014virtual} in which the power electronics interface (PEI) is mimicking the behavior of the SGs.
However the implementation of virtual inertia in the literature varies based on the desired level of of model complexity and aplication, the underlying concept is similar among various topologies. Using a detailed mathematical model which represents the dynamics of SG or simpifying the model by 
using only the swing equation are the the two main solutions for implementing virtual inertia \cite{tamrakar2017virtual,vassilakis2013battery}.
For example to represent the same dynamics as SGs, synchroverters are introduced in \cite{zhong2010synchronverters} for inverter-based DGs. Hence as an advantage, traditional operation of power system can be continued without 
significant canges in operation structure \cite{zhong2016virtual}.  
Similar to synchroverter approach, Ise lab is another topology in literaure to implement virtual inertia. In this method, the contro loop solve the swing equation in 
each cycle to emulate inertia, instead of using a full detail model of SG \cite{sakimoto2011stabilization}.
Other than different topologies VSG application has been illuterated extensively in the literature.
In \cite{liu2016enhanced} an enhanced VSG control is proposed in which by adjusting the virtual stator reactance, active power oscillations during transient states is improved. Furturmore, using inverse voltage droop control and ac bus voltage estimation, an accurate reactive power sharing is achieved.  
In \cite{liu2015comparison}, the dynamic characteristic of simple droop control and VSG are studied through deriving small signal equations in both islanded and grid connected mode. Then the proposed inertial droop control by the author, inherits the advantage of droop and provides inertial support for the system. Voltage angle deviations (VADs) of generators with respect to the
angle of the center of inertia are defined in \cite{alipoor2016stability} as a tool for transient
stability assessment of the multi-VSG micro-grid. To have a smooth transition during disturbances and keeping VADs within a specific range, VSG parameters are tuned 
using particle swarm optimization. 
In \cite{wu2016small} the detailed parameter designs of VSG is proposed and also the conditions to decouple active and reactive power loops are given. For avoiding VSG output voltage distortions the author indicates indicates that the bandwidth of the power loop should be very smaller than twice the line frequency  
In \cite{wu2016virtual} to enhance the inertial response of DC micro-grids and stabilize the DC bus voltage fluctuations, the author proposed virtual inertia strategy for DC micro-grids through bidirectional grid-connected converters. A fuzzy-secondary-controller-based VSG control scheme is proposed in \cite{andalib2018fuzzy} for voltage and frequency regulation in micro-grids. \cite{shi2017low} proposed a low voltage ride through control strategy for VSG control scheme with providing reactive power support under grid fault. The solution strategy for VSG working under unbalanced voltage conditions is discussed in \cite{zheng2017comprehensive}. 
A new VSG control in presented in \cite{zhao2017multi} with capability to avoid harmonic interference and accurate control vector orientation process.

This paper is organized as follows. Section \RomanNumeralCaps{2} introduces the proposed
power grid with DERs. Section \RomanNumeralCaps{3} discusses impacts of DERs on frequency while the solution for this impact is discussed in Section \RomanNumeralCaps{4}. Finally, simulation and experimental results and conclusion are shown in  and Section \RomanNumeralCaps{5} and Section \RomanNumeralCaps{6} receptively.
\begin{figure}[!t]\label{pic12}
	\centering{\includegraphics[scale=0.6]{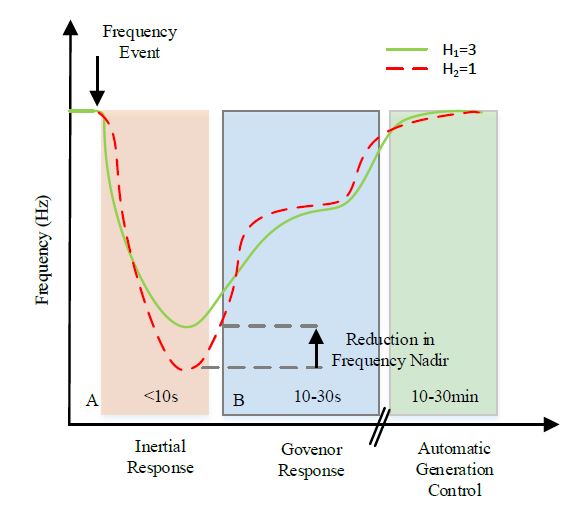}}
	\caption{Multiple time-frame frequency response in a power system following a frequency event. \label{fig12}}
\end{figure}

	\begin{figure*}[!t]\label{pic11}
	\centering{\includegraphics[scale=0.6]{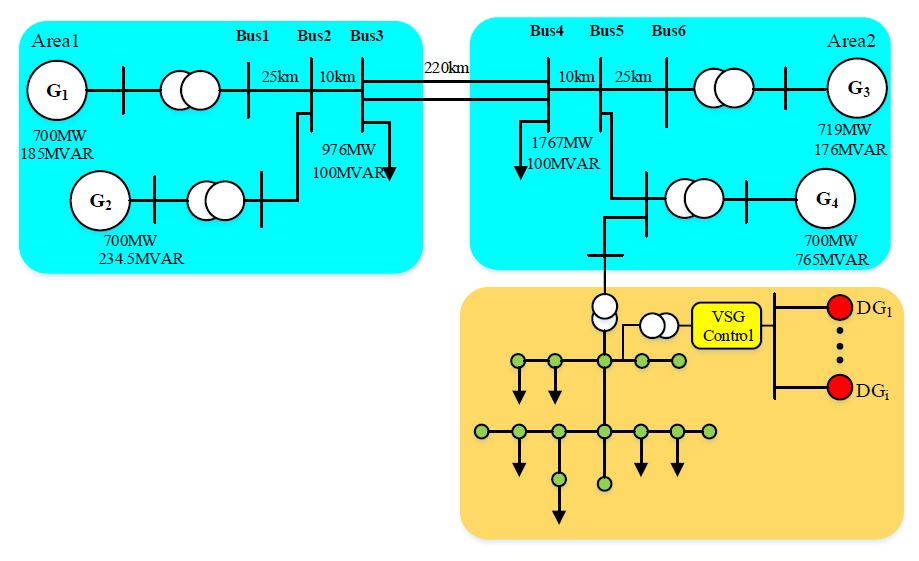}}
	\caption{T\&D combined system with VSG technologies. \label{fig11}}
\end{figure*}

	\begin{figure*}[!t]\label{pic1}
		\centering{\includegraphics[scale=0.8]{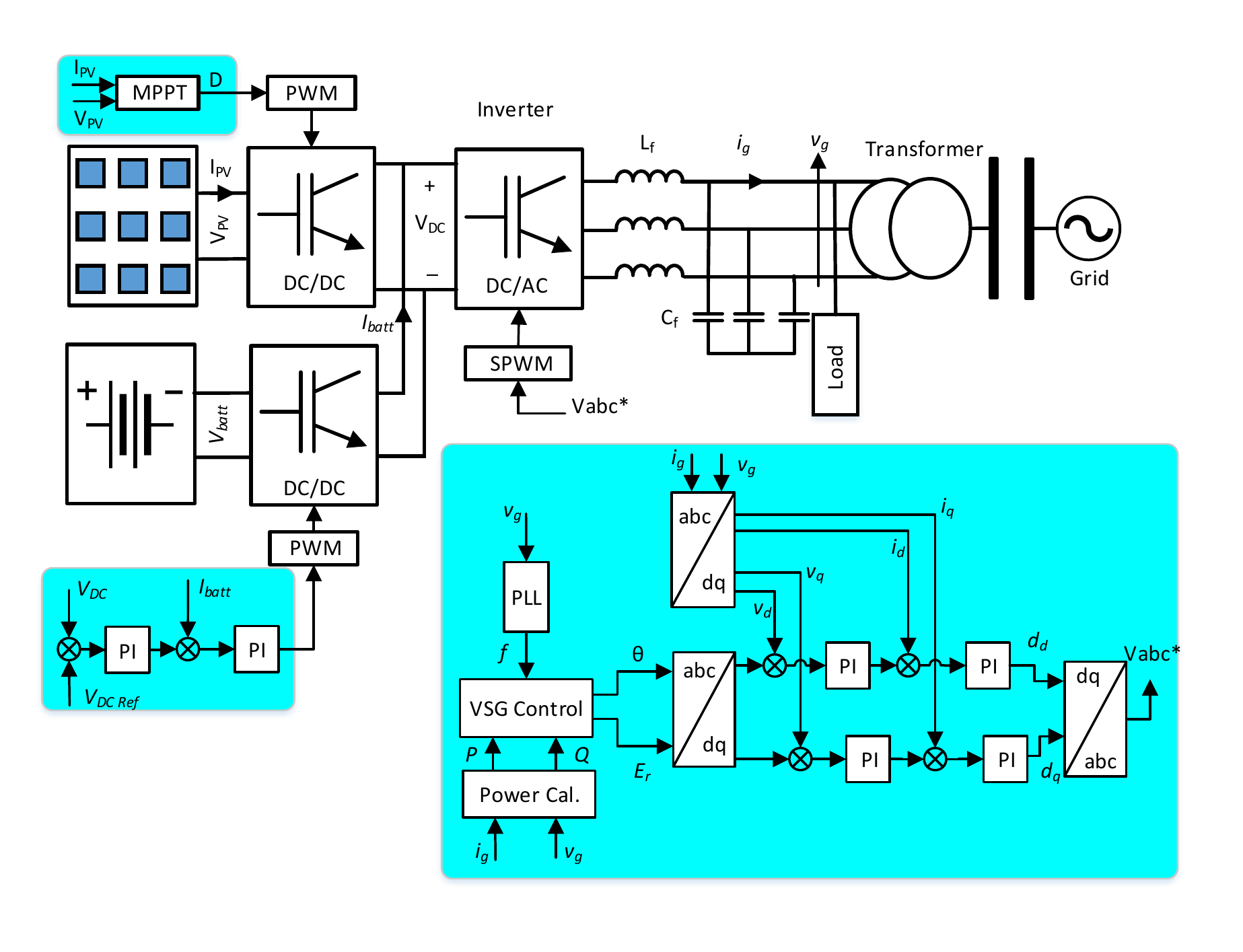}}
		\caption{General schematic representation of the proposed VSG controller. \label{fig1}}
	\end{figure*}
	
	\section{transmission and distribution system}  \label{sec2}
	
Fig. \ref{fig11} shows the sample power grid with generation, transmission and distribution system. The Kundur's two-Area system with parameters taken from \cite{kundur1994power} which is comprised of four synchronous generators(two in each area) that are boosting up with transformers and connected through transmission lines, form generation and transmission parts in this example. The areas are connected to each other through a tie line. IEEE 13 node test feeder \cite{kersting1991radial} is used as distribution system. The DGs such as PV and battery energy storage unit (BES) are connected to distribution system using a voltage source inverter (VSI).

	PV system includes PV arrays and a unidirectional boost DC-DC converter which is working under perturb and observe (P\&O) maximum power point tracking (MPPT) control . BES includes batteries and a bidirectional boost DC-DC converter controlled by multi-loop voltage and current control. The outer loop controls voltage and inner loop controls current through proportional integral (PI) controllers.
	The BES and PV unit are connected in parallel and form the DC link.
	
	\section{impact of der on frequency}
	Since the power electronic interfaces used in DGs has no rotating mass and damping, the inertial constant in the micro-grid is reduced which results in an increase in the rate of change of frequency (ROCOF) and may lead to load shedding even under small disturbances in the system. Fig. \ref{fig12} shows the frequency curve of system with different amount of inertia in presence of a contingency. For the control and stability of these small scale power grids, a hierarchical control including primary, secondary and tertiary control is introduced, similar to conventional grids. Droop control for voltage source inverters as an example for primary frequency control is discussed in \cite{rocabert2012control}, provide barely any inertia/damping support for the grid.
	
		\section{VSG control}\label{sec3}
		
\begin{figure}[!t]\label{pic2}
		\centering{\includegraphics[scale=0.8]{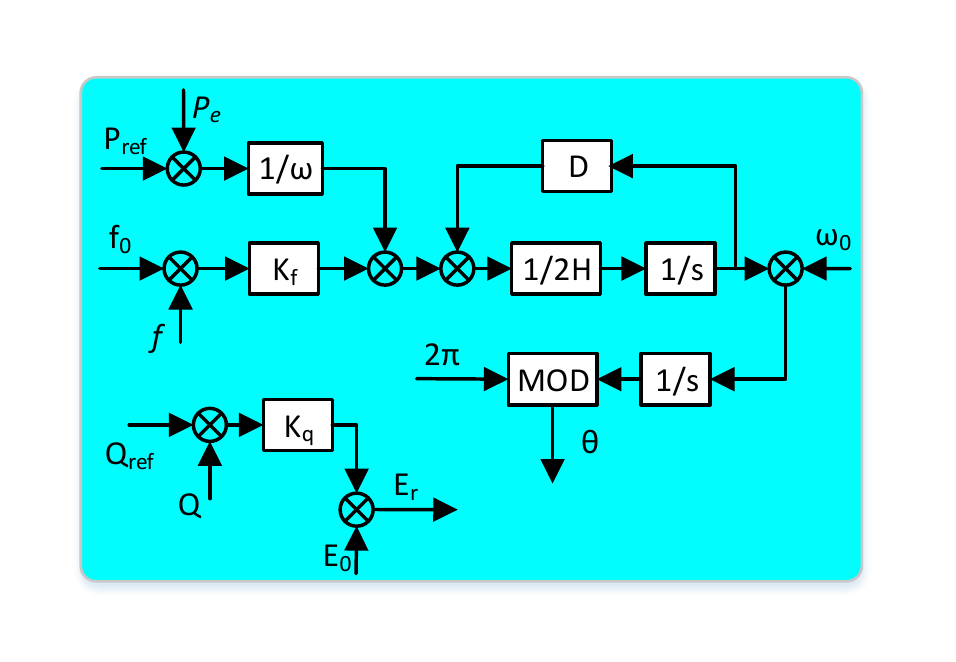}}
		\caption{General schematic representation of the proposed VSG controller. \label{fig2}}
	\end{figure}

	Without any mechanical rotational part, inverters have a high response speed compared to the conventional rotational machines \cite{alipoor2013distributed}. Virtual inertia concept is introduced as a solution to overcome this limitation \cite{shintai2014oscillation}.   
	By emulating the mechanical equation of a real synchronous generator into the inverter, similar behavior can be assumed during normal operation of the system and frequency disturbances for example the time that there is a sudden change (increase or decrease) in active power. Utilizing VSG algorithm, synchronized active power can be injected from the PV to the grid to stabilize the frequency \cite{fathi2018robust,sakimoto2012stabilization}. In this paper, during normal operation of the system (rated frequency and voltage), the perturb and observe method (P\&O) sets the active power reference $P{ref}$ by measuring voltage and current of the PV \cite{zhang2012review}. This active power is controlled in two stages as shown in Fig. \ref{fig2}. At the first stage, primary frequency is implemented in the same way as a SG. In the second stage virtual inertia and damping are added to complete the loop. The result is a reference angle that will be fed into park transform \cite{du2013modeling}.     \\
	VSG control can be divided into two sections. First, the mechanical swing equation needs to be emulated and solved numerically. Then the results are used as a reference to control the voltage and current of the inverter.
		
		\subsection{P-F control}
		Mechanical equation of a SG assuming that the rotor is a rigid body can be described as 
		
		\begin{align}\label{eq3}
		\begin{cases}
		\frac{d\theta}{dt}=\omega   \\
		2H\frac{d\omega}{dt}=T_{m}-T_{e}-D\Delta\omega
		\end{cases} 
		\end{align}
		
		where $H$ is the inertia constant in p.u. derived from (\ref{swing}), $S_{base}$ is the base power of the machine, $\omega$ is the angular frequency of the SG and $\omega _{0}$ the rated angular frequency \cite{zeng2015mathematical}.
		
		 	\begin{equation} \label{swing}
		 J = 2H\frac{{S_{base}}}{\omega _{0}^{2}} 
		 \end{equation}
		 
		 $T_{m}$ and $T_{e}$ are the mechanical output torques of the prime mover and the electromagnetic torque of the SG respectively and can be calculated using  (\ref{eq4}) :
		 
		\begin{equation}\label{eq4}
		\begin{cases}
		T_{m}=k_{f}(f_{0}-f)+\frac{P_{ref}}{\omega} \\
		T_{e}=\frac{P_{e}}{\omega}
		\end{cases} 
		\end{equation}
		
		in which $P_{ref}$ is the rated active power and $P_{e}$ is the output power of the DG. The primary frequency control method and damping coefficient are working here as in the similar to a real SG and are achieved through a proportional cycle in which $k_{f}$ is the droop coefficient and $D$ is called the damping coefficient.
	for typical synchronous machines $H$ varies between 2 and 10 s \cite{hirase2016analysis}. The VSG response at a specific output power and voltage is determined by parameters of its second order differential equation which are the real part of its eigenvalues $\sigma_{i}$ and the damping ratio $\xi_{i}$.These parameters are related to $J$ and $D$ directly through the following equation set
	
\begin{equation} 
\begin{split}
\sigma _{i}=&- \frac {D_{i}}{{2{J_{i}}{\omega _{s}}}}    \\
{\omega _{ni}}=&\sqrt {\frac {{{P_{\max i}}\cos \left ({ {{\theta _{ig}}} }\right )}}{{J_{i}{\omega _{s}}}}}  \\
{\xi _{i}}=&\frac {{ - {\sigma _{i}}}}{{{\omega _{ni}}}}
\end{split}
\end{equation}

where $P_{maxi}$ is the maximum transferable power from the VSG bus to the grid, $\theta_{ig}$ is the voltage angle of the  VSG with respect to the grid and $\omega_{ni}$ is the undamped natural frequency of the VSG. At any working conditions, the parameters corresponding to the desired system response can be achieved by tuning $J$ and $D$ \cite{li2015coherency}.
		
		\subsection{Q-E regulation}
		
		controlling voltage is achieved by regulating the reactive power as (\ref{eq5}) 
		
		\begin{equation}\label{eq5}
		E_{r}=E_{0}+k_{q}(Q_{ref}-Q)
		\end{equation}
		
		where $E_{r}$ is the reference voltage, $k_{q}$ is reactive power droop coefficient and $E_{0}$ is the nominated voltage amplitude. 
		
		\subsection{V/I control}
		This control loop features a conventional outer voltage and inner current loop. Its primary function is regulating the output voltage with no steady state error while quickening the dynamic response of the current loop to strengthen the ability of inverter control. This can be achieved through the outer voltage and inner current control loop. The calculated reference voltage $E_{r}$ using (\ref{eq5}) is set as the reference for the outer voltage loop. Since the control is achieved in constant reference frame, $E_{r}$ is  transformed from synchronous to dq reference frame using Park transformation given in (\ref{eq7}) with the calculated angle in (\ref{eq3}) as the input angle for the transformation. Then a PI controller is tuned to track the reference voltage and current. 
		
		\begin{equation}\label{eq6}
		E_r=\begin{bmatrix}
		E_{ar}\\
		E_{br}\\
		E_{cr}
		\end{bmatrix}=
		\begin{bmatrix}
		E\sin\omega t\\
		E\sin\omega(t-\frac{2\pi}{3}) \\
		E\sin\omega(t+\frac{2\pi}{3}) 
		\end{bmatrix}
		\end{equation}

		\begin{equation}\label{eq7}
		\begin{bmatrix}
		V_{d}^\ast\\
		V_{q}^\ast\\
		V_{0}^\ast
		\end{bmatrix}=\sqrt{\frac{2}{3}}
		\begin{bmatrix}
		\cos\gamma & \cos(\gamma-\frac{2\pi}{3}) & \cos(\gamma+\frac{2\pi}{3})\\
		\sin\gamma & \sin(\gamma-\frac{2\pi}{3}) & \sin(\gamma+\frac{2\pi}{3})\\
		\frac{1}{\sqrt2} & \frac{1}{\sqrt2} & \frac{1}{\sqrt2}
		\end{bmatrix}
		\begin{bmatrix}
		E_{ar}\\
		E_{br}\\
		E_{cr}
		\end{bmatrix}
		\end{equation}
		
		The current of the converter is also controlled similarly using an inner PI controller.

	\section{Simulations and results}~\label{simulations}
		
\begin{figure}[!t]\label{pic9}
	\centering{\includegraphics[scale=.7 ]{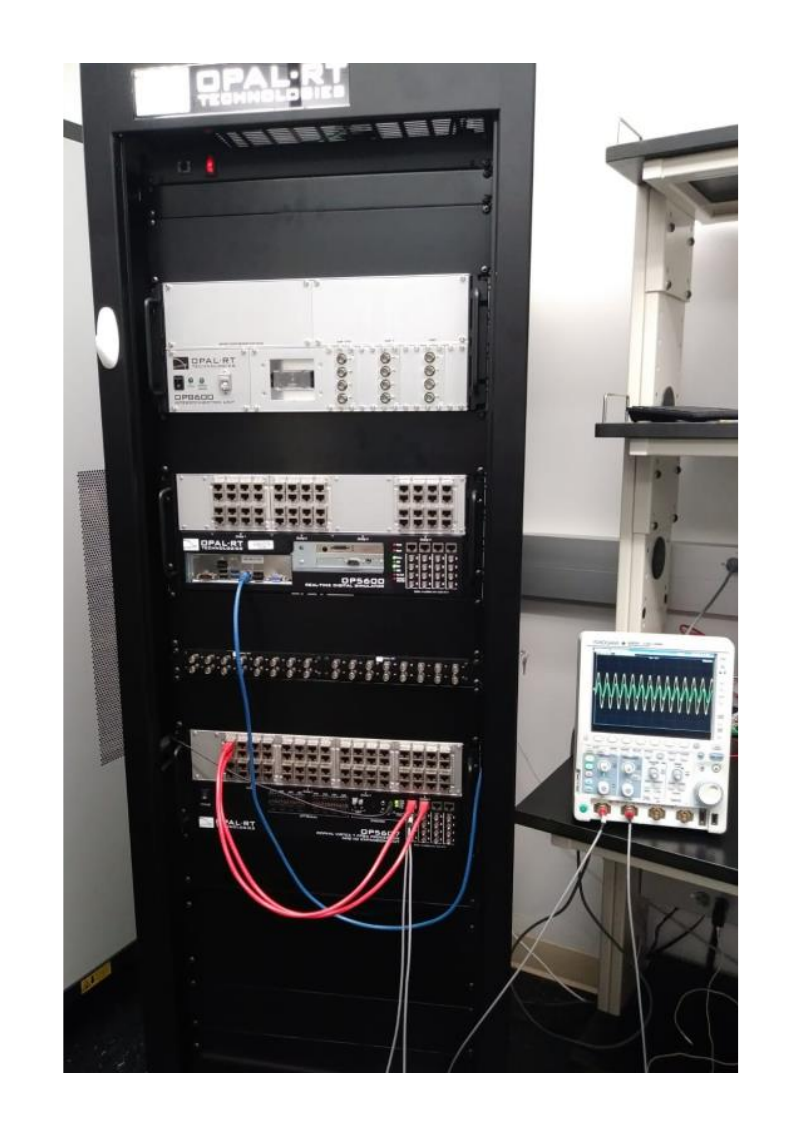}}
	\caption{Opal RT simulator.\label{fig16}}
\end{figure}

As illustrated in Fig. \ref{fig1}, the proposed VSG 
 is tested for different case studies such as normal condition and sudden changes in loads. The converter and control is simulated using Matlab/Simulink SimPowerSystems toolbox. Real time results are achieved using Opal Rt OP5600 and OP5607 Virtex7 FPGA simulator which is shown in Fig. \ref{fig16}. 
The converter switches are simulate in FPGA due to its capability of working with smaller sampling time with $T_{s}=1 \mu s$. The  rest of simulation are implemented in simulator target with $T_{s}=10 \mu s$. The sampling time for both the simulator and FPGA is chosen based on their desired performance and also complexity of the control scheme.


	
	\subsection{Frequency Regulation}
	
\begin{figure}[!t]\label{pic5}
		\centering{\includegraphics[width=3.25in]{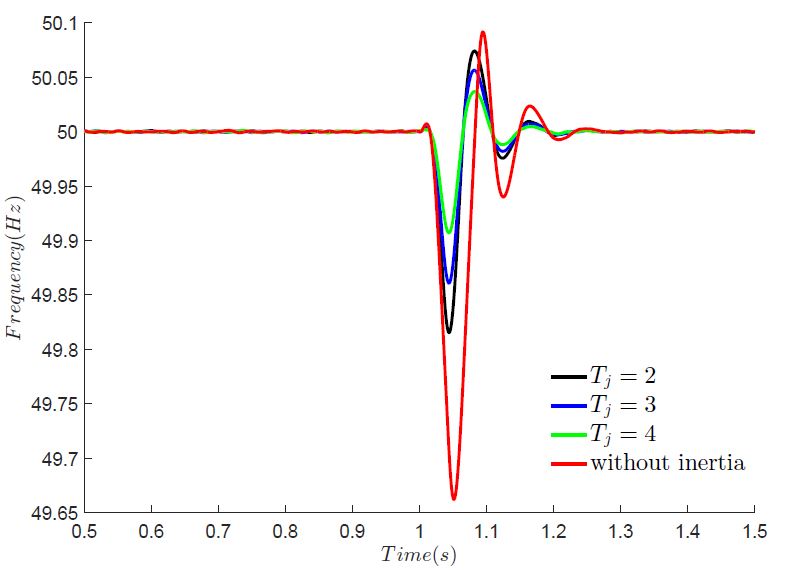}}
		\caption{Frequency deviation of the proposed system under different value of $T_{j}$ and 0.2 p.u. step load.\label{fig5}}
	\end{figure}

\begin{figure}[!t]\label{pic7}
	\centering{\includegraphics[width=3.25in]{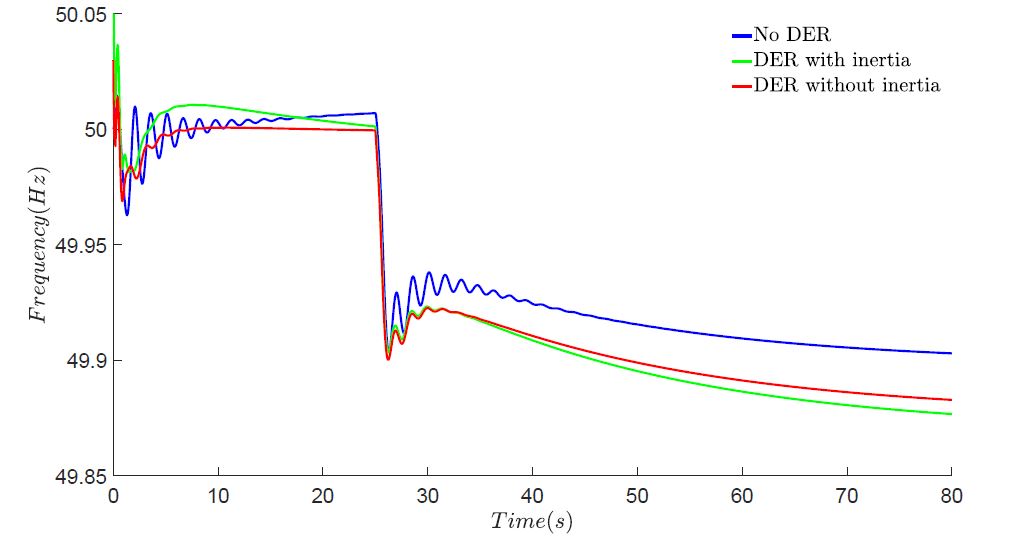}}
	\caption{System frequency with no DER, with DER and no inertia emulation and with DER and inertia emulation respectively at 200MW.\label{fig7}}
\end{figure}

\begin{figure}[!t]\label{pic8}
		\centering{\includegraphics[width=3.25in]{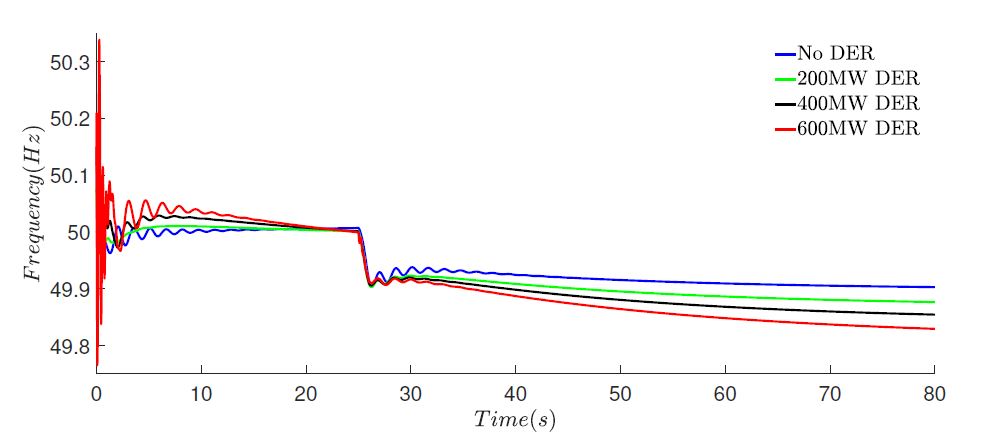}}
		\caption{Effect of DER with VSG on BES frequency.\label{fig8}}
	\end{figure}

	 First the system is working under normal operation condition. The generated power by PV under MPPT is used to feed the load at the distribution system and rest of that is sent to the grid to feed the grid loads. At this scenario, the frequency behavior of the system is tested under different inertia constants ($T_{j}=2H$) with 0.2 p.u increase in active load at $t=1 s$ in which the total load is still less than the total generated MPPT power. The frequency nadir is shown and compared in Fig. \ref{fig5}. The smaller inertia constant $T_{j}$, the bigger frequency nadir and the deepest nadir is expected for the case when there is no inertia in system. The bigger frequency nadir results in a higher risk for the system that can cause it to disconnect from the main grid due to triggering ROCOF relays. In next scenario, the increase in load is equal to the total generation of the PV.\\
	In Fig. \ref{fig7} the system frequency performance is compared with the addition of $200MW$ DER and state-of-the-art power electronics inverters that doesn't emulate inertia versus DERs with the proposed VSG control. The results shows the frequency nadir decreased with the addition of the virtual inertia.
	Fig. \ref{fig8} depicts the system frequency response corresponding to the different DER penetration levels. It can be seen that the VSG is emulating the generator inertia and therefore even with a big increase in the DER penetration level, the frequency nadir is almost the same as with the generators only.
	
\section{conclusion}

DERs have multiple negative impacts on the bulk power systems which have been addressed in the paper. In between, lower or zero inertia is one of their major aspects that will affect the stability of the whole power grid and may lead into unwanted load shedding. The solution for adding inertia to DERs by mimicking synchronous generation behavior is introduced. By using the proposed seamless control framework, a simple control strategy is implemented to operate under normal and faulty grid conditions. Several experiments have been conducted to verify the performance of the proposed system.

	\section*{Acknowledgments}
	This project was funded in-part by the University of Texas at San
	Antonio, Office of the Vice President for Research.

	\bibliographystyle{IEEEtran}	\bibliography{bibl}
	\vspace{-0.48cm}

\end{document}